# Machine learning for predicting control landscape maps of quantum molecular dynamics: Laser-induced three-dimensional alignment of asymmetric top molecules

Tomotaro Namba[1)] and Yukiyoshi Ohtsuki*

Department of Chemistry, Graduate School of Science, Tohoku University

6-3 Aramaki Aza-Aoba, Aoba-ku, Sendai 980-8578, Japan

1) present address

Nuclear Science and Engineering Center, Japan Atomic Energy Agency

2-4 Shirakata, Tokai-mura, Naka-gun, Ibaraki 319-1195, Japan

* Corresponding author: yukiyoshi.ohtsuki.d2@tohoku.ac.jp

Tel/fax: +81-22-795-7725

(August 30, 2024)

**Abstract**

Machine learning for predicting control landscape maps of full quantum molecular dynamics is examined through a case study of the laser-induced three-dimensional (3D) alignment of asymmetric top molecules, an essential technique for observing and/or manipulating molecular dynamics in a molecule-fixed frame. We consider the "prolate-type" asymmetric top molecules with the asymmetry parameters $-1 < \kappa < 0$ and the $C_{2v}$ symmetry in the low-temperature limiting case, which are aligned by using mutually orthogonal linearly polarized double laser pulses. The landscape map for each molecule consists of 6000 pixels, each pixel of which represents the maximum degree of alignment achieved by each set of control parameters. After examining ways to deal with the markedly different molecular parameters in a unified manner for suitably training a convolutional neural network (CNN) model, we train the CNN model by using 55 training sample molecules to predict the control landscape maps of 35 test sample molecules with reasonably high accuracy. As the predicted landscape maps provide a big picture of the alignment control, we show, for example, that the double pulse control scheme is especially effective for a molecule having a polarizability component that is much larger in value than the other two components.

# I. INTRODUCTION

Our capability to simulate full quantum molecular dynamics is severely limited by quantum correlations due to the many-body interactions even when the initial states are localized because the quantum entanglements spread over the entire systems with time. Machine learning (ML) approaches have attracted immense attention as a means to overcome these difficulties [1–3]. It would be convenient to classify the ML approaches into three categories. In the first category, ML approaches are applied to determine molecular properties such as potential energy surfaces and forces while avoiding computationally expensive electronic structure calculations [4–6] typically used in model-based simulations. In the second category, variational representations of quantum states such as neural-network quantum states are proposed on the basis of the ansatz that tractable subspaces of total Hilbert spaces may represent many-body wavefunctions [7–10]. Some studies [10] consider the time-dependent expectation values of the set of observables instead of the wavefunctions because they can be straightforwardly connected to the experimental measurements. ML approaches in the second category are used to predict the excitation energy transfer , e.g., in the Fenna–Matthews–Olson (FMO) complex [11–15], the optimal control design [16,17], the landscapes of the cost functions [18] and so on.

There is third category of ML approaches that provide computationally effective tools for quantum molecular dynamics simulation. In this category, we simulate the dynamics by exactly solving the Schrödinger equation at the expense of high computational costs. Vast computer resources, albeit not practically feasible, are required to simulate the dynamics of various molecules in a unified manner, e.g., to find the mechanisms and/or the rules of thumb for valid control across the molecules. The purpose of the present study is to consider one example through a case study of laser-induced three-dimensional (3D) molecular alignment [19–22]. To be specific, we examine ML approaches for predicting control landscape maps of the 3D alignment control, i.e., the full quantum molecular dynamics. Alignment control is essential to realize further control of molecular dynamics and/or measurements in the molecule-fixed frame.

To achieve molecular alignment in the field-free condition, we often utilize polarizability interactions induced by mildly intense non-resonant laser pulses, taking advantage of the controlled intense electric fields of the laser pulses [23]. Because it is obvious that a higher degree of 3D alignment is better for applications, several control schemes for effective 3D alignment of asymmetric molecules have been proposed [24–34]. One control scheme proposes the use of a train of elliptically polarized laser pulses with optimized ellipticities, which are suitably adjusted to the polarizability components of the target molecules [24,25]. In another control scheme, a pair of mutually orthogonal linearly polarized laser pulses is adopted, in which the first pulse 1D aligns the most polarized molecular axis along the polarization vector and the second orthogonal pulse interacts with the other molecular axes at the timing when the torque exerted on the first molecular axis is minimized [26,27]. In our previous study [33], we considered the laser-induced 3D alignment of $SO_2$, a typical molecule in proof-of-principle experiments and theoretical studies, and derived the mutually orthogonal pulse train as the optimal solution by using the optimal control theory. We also showed that the optimal pulse train can be approximated by the mutually orthogonal double pulses while reasonably maintaining the effectiveness of the 3D alignment control. All of the above-mentioned studies focused on their own "specific" molecules, intending to clarify some general features of the alignment control through specific examples. It would be, thus, natural to consider the 3D alignment control of a certain class of asymmetric top molecules in a unified manner by resolving computational-cost issues with, e.g., convolutional neural network (CNN) models.

In the present study, assuming alignment control with a pair of mutually orthogonal linearly polarized laser pulses, we apply the control scheme to "prolate-type" asymmetric top molecules with $C_{2v}$ symmetry with the aim of examining how much we could increase their degrees of 3D alignment in the control scheme, clarifying the molecule-dependent control mechanisms, and so on. For these purposes, we focus on control landscape maps [35–37] defined by the maximum degrees of 3D alignment as a function of control parameters [37]. Assuming induced-dipole interactions, we specify the asymmetric top molecules by using sets of three rotational constants and polarizability tensors, the values of which are widely distributed over 2-3 orders of magnitude depending on the molecule. In addition, the asymmetric top molecules are not uniformly distributed in the "molecular space" in which the CNN model is trained. Therefore, the training efficiency is severely reduced and CNN model performance deteriorates. Our purpose is to examine how to overcome these difficulties by actively utilizing dimensionless molecular and control parameters and by introducing "artificial" molecules. We show that the present CNN model predicts the control landscape maps with reasonable accuracy, whereby the general features of the alignment control can be semi-quantitatively examined.

This manuscript is organized as follows. In Sec. II, we briefly summarize a numerical method to solve the laser-induced rotational dynamics and define the control landscape maps in a general way. After introducing a class of asymmetric top molecules considered in the present study, we explain how to prepare samples suitable for CNN model training. In Sec. III, we show and discuss the results, and in Sec. IV, we summarize the present study.

## II. Theory

### A. Time evolution of rotational wave packets and control landscape maps

We consider asymmetric top molecules interacting with pairs of mutually orthogonal linearly polarized non-resonant laser pulses, the polarization vectors of which are assumed to be parallel to the space-fixed *X*- and *Y*-axes. We introduce the unit vectors $\boldsymbol{e}_X$, $\boldsymbol{e}_Y$, and $\boldsymbol{e}_Z$ ($\boldsymbol{e}_a$, $e_b$, and $e_c$) of the space-fixed *XYZ*-axes (the molecule-fixed principal axes of inertia). The time evolution of the rotational dynamics is described by the wave packet $|\psi(t)\rangle$ that obeys the Schrödinger equation

$$i\hbar \frac{\partial}{\partial t}|\psi(t)\rangle = H(t)|\psi(t)\rangle = [H_0 + V(t)]|\psi(t)\rangle, \tag{1}$$

with the initial condition $|\psi(t_0)\rangle = |\psi_0\rangle$. In Eq. (1), $H_0$ and $V(t)$ are the field-free Hamiltonian and the laser-molecule interaction, respectively. If we introduce the *dimensionless* angular momentum operator $\boldsymbol{J}$, which is the angular momentum operator divided by $\hbar$, the field-free Hamiltonian [38] is expressed as

$$H_0 = h(AJ_a^2 + BJ_b^2 + CJ_c^2), \tag{2}$$

where the components of the dimensionless angular momentum operator are defined by $J_a = \boldsymbol{e}_a \cdot \boldsymbol{J}$, etc. In Eq. (2), the rotational constants are defined by $A = \hbar/4\pi I_{aa}$, $B = \hbar/4\pi I_{bb}$, and $C = \hbar/4\pi I_{cc}$ ($A > B > C$) with $I_{aa}$ being the *a* component of the principal moment of inertia, etc.

As we assume mildly intense non-resonant laser pulses, we consider the lowest-order induced-dipole interaction and take the cycle average over the optical frequency. We then have the interaction

$$V(t) = -\frac{1}{4}\alpha_{XX}[\mathcal{E}_X(t - t_X^0)]^2 - \frac{1}{4}\alpha_{YY}[\mathcal{E}_Y(t - t_Y^0)]^2, \tag{3}$$

where $\alpha_{XX}$ ($\alpha_{YY}$) is the space-fixed $XX$ ($YY$) component of the polarizability tensor. In Eq. (3), $\mathcal{E}_X(t-t_X^0)$ and $\mathcal{E}_Y(t-t_Y^0)$ are the envelope functions of the laser pulses linearly polarized along the space-fixed $X$-axis and $Y$-axis, respectively. Here, we assume the temporal peak positions $t_X^0$ and $t_Y^0$, which define the time delay $\tau = t_Y^0 - t_X^0 \, (\geq 0)$. According to the temporal widths of the laser pulses and the grid size associated with $\tau$, which will be explained in Sec. IIC, we can safely ignore the temporal overlap between $\mathcal{E}_X(t-t_X^0)$ and $\mathcal{E}_Y(t-t_Y^0)$ when $\tau > 0$. In the case of $\tau = 0$, we assume the fixed value of $\pi/2$ for the phase difference in the optical frequency between the $X$ and $Y$ polarized laser pulses as we are not interested in the control with elliptically polarized laser pulses in the present study. Because of this, the interaction potential is always given by Eq. (3) independent of the value of $\tau \geq 0$. The total pulse fluence $F$ is given by

$$F = \frac{1}{2}\varepsilon_0 c \int_{-\infty}^{\infty} dt \left\{ [\mathcal{E}_X(t)]^2 + [\mathcal{E}_Y(t)]^2 \right\} \equiv F_X + F_Y \equiv rF + (1-r)F, \qquad (4)$$

where $\varepsilon_0$ and $c$ are the dielectric constant in vacuum and the speed of light, respectively. We introduce the ratio $r$ that determines how to divide the total pulse fluence $F$ into the fluence of the $X$-polarized laser pulse $F_X$ and that of the $Y$-polarized laser pulse $F_Y$. We focus on $\tau = t_Y^0 - t_X^0$ and $r$ as the control parameters, assuming that the temporal widths of the laser pulses are fixed.

It is necessary to deal with various asymmetric top molecules, the rotational constants of which are distributed in the range of 2 to 3 orders of magnitude in value. In order to treat them in a unified manner, we measure the energy in units of the rotational constant $hA$. The Schrödinger equation is expressed in the dimensionless form as

$$i\frac{\partial}{\partial t}|\psi(t)\rangle = [J_a^2 + \tilde{B}J_b^2 + \tilde{C}J_c^2 + \tilde{V}(t)]|\psi(t)\rangle, \tag{5}$$

where $\tilde{B} = B/A$, $\tilde{C} = C/A$ ($0 < \tilde{C} < \tilde{B} < 1$), and $\tilde{V}(t) = V(t)/hA$. Although we use the same notation "$t$" as that in Eq. (1), we introduce the dimensionless time $t$ in Eq. (5), which is defined by multiplying the original time by $2\pi A$. Under given sets of laser pulses $\mathcal{E}_X(t - t_X^0)$ and $\mathcal{E}_Y(t - t_Y^0)$, we numerically integrate Eq. (5) during the period $[t_0, t_f]$, where $t_f$ specifies the final time. We expand the rotational wave packet $|\psi(t)\rangle$ in terms of the eigenfunctions of a symmetric top $\{|JKM\rangle\}$, which satisfy the eigenvalue equations $\boldsymbol{J}^2|JKM\rangle = J(J+1)|JKM\rangle$, $J_c|JKM\rangle = K|JKM\rangle$ ($K = -J, -J+1, \cdots, J$) with $J_c = \boldsymbol{e}_c \cdot \boldsymbol{J}$, and $J_Z|JKM\rangle = M|JKM\rangle$ ($M = -J, -J+1, \cdots, J$) with $J_Z = \boldsymbol{e}_Z \cdot \boldsymbol{J}$. The details of the numerical integration are summarized in Appendix A. For simplicity, we ignore the nuclear spin statistics.

Next, we define the degrees of 3D alignment as a function of the control parameters, i.e., the time delay $\tau = t_Y^0 - t_X^0$ [Eq. (3)] and the ratio $r$ [Eq. (4)]. To be specific, we calculate the expectation values of all possible combinations, i.e., a total of six combinations of the direction cosines, which are defined by

$$\varphi_{abc}(\tau,r;t) = \frac{1}{3}\langle\psi(t)|[(\boldsymbol{e}_a\cdot\boldsymbol{e}_X)^2 + (\boldsymbol{e}_b\cdot\boldsymbol{e}_Y)^2 + (\boldsymbol{e}_c\cdot\boldsymbol{e}_Z)^2]|\psi(t)\rangle,$$

$$\cdots \quad (6)$$

$$\varphi_{cba}(\tau,r;t) = \frac{1}{3}\langle\psi(t)|[(\boldsymbol{e}_c\cdot\boldsymbol{e}_X)^2 + (\boldsymbol{e}_b\cdot\boldsymbol{e}_Y)^2 + (\boldsymbol{e}_a\cdot\boldsymbol{e}_Z)^2]|\psi(t)\rangle,$$

after the temporal peak of the second pulse. During $t \in [t_Y^0, t_f]$, we search the maximum value for a given set of control parameters $\tau$ and $r$, which is denoted by $\Phi(\tau,r)$. The contour map of $\Phi(\tau,r)$ as a function of $\tau$ and $r$ defines the control landscape map in the present study. If the landscape map is composed of, for example, 6000 $\Phi(\tau,r)$ values, we have to solve Eqs. (5) and (6) 6000 times for each molecule. The landscape map contains abundant information on the 3D alignment control, but its construction requires a high computational cost. This has motivated us to adopt the ML approaches to reduce the computational cost.

**B. Asymmetric top molecules considered in the present study**

Because of the issue of the high computational cost mentioned above, we focus on some specific classes of asymmetric top molecules to reduce the number of training samples. We introduce the first criterion associated with the asymmetry parameter $\kappa = (2B-A-C)/(A-C) = (2\tilde{B}-1-\tilde{C})/(1-\tilde{C})$, which takes a value between $-1$ for a prolate top ($B=C$) and $+1$ for an oblate top ($A=B$) [38]. We consider the prolate-type asymmetric top molecules with $-1<\kappa<0$. The second criterion is about the principal axes. The principal axes of inertia characterize the free rotational dynamics, whereas those of polarizability characterize the "torques" received from the laser pulses. Without significant loss of generality, we consider the molecules with $C_{2v}$ symmetry so that the two principal-axis systems become the same. Because of the $C_{2v}$ symmetry, the polarizability interaction in Eq. (3) is reduced to

$$V(t) = -\frac{1}{4}\Big[\Delta\alpha_{ac}(\boldsymbol{e}_a \cdot \boldsymbol{e}_X)^2 + \Delta\alpha_{bc}(\boldsymbol{e}_b \cdot \boldsymbol{e}_X)^2\Big](\mathcal{E}_X^0)^2 \varphi_X(t - t_X^0)$$
$$-\frac{1}{4}\Big[\Delta\alpha_{ac}(\boldsymbol{e}_a \cdot \boldsymbol{e}_Y)^2 + \Delta\alpha_{bc}(\boldsymbol{e}_b \cdot \boldsymbol{e}_Y)^2\Big](\mathcal{E}_Y^0)^2 \varphi_Y(t - t_Y^0),$$
(7)

where $\Delta\alpha_{ac} = \alpha_{aa} - \alpha_{cc}$ and $\Delta\alpha_{bc} = \alpha_{bb} - \alpha_{cc}$. In Eq. (7), the squares of the laser pulses are assumed to be expressed as $[\mathcal{E}_X(t - t_X^0)]^2 = (\mathcal{E}_X^0)^2 \varphi_X(t - t_X^0)$ and $[\mathcal{E}_Y(t - t_Y^0)]^2 = (\mathcal{E}_Y^0)^2 \varphi_Y(t - t_Y^0)$ for convenience, where the normalized non-negative functions $\varphi_X(t - t_X^0)$ and $\varphi_Y(t - t_Y^0)$ representing the intensity envelope functions with the temporal peak positions $t_X^0$ and $t_Y^0$, respectively, are assumed to be given functions of time. From Eq. (7), the values of $(\mathcal{E}_X^0)^2$ and $(\mathcal{E}_Y^0)^2$ are specified by the total pulse fluence $F$ and the ratio $r$ [see Eq. (4)]. Because of the first criterion $-1 < \kappa < 0$, most of the molecules are characterized by $-1 < \kappa_\alpha < 0$ if we introduce the asymmetry parameter associated with the polarizability defined by $\kappa_\alpha = (2\alpha_{bb} - \alpha_{aa} - \alpha_{cc})/(\alpha_{aa} - \alpha_{cc})$. We thus remove "exceptional molecules", the $\kappa_\alpha$ values of which are not within the range of $-1 < \kappa_\alpha < 0$. This "weak" restriction is referred to as the third criterion.

We restrict ourselves to the electrically neutral molecules in the electronic ground states with spin singlet states and take the asymmetric top molecules that satisfy the above-mentioned three criteria from the NIST database [39] except for SO$_2$. By using the coordinates provided by the NIST database as the initial molecular structures, we re-optimize the structures by RHF/6-311G (d, p) with the Gaussian 16 program package and obtain the rotational constants. We then have 65 molecules remaining because we remove molecules that lose the $C_{2v}$ symmetry after our optimization calculations. We calculate the polarizabilities by adopting CCSD/cc-pVDZ. For $SO_2$, the parameters are taken from Refs. [40] and [41].

The rotational dynamics considered here is expressed in terms of $\tilde{B}$, $\tilde{C}$, $\Delta\alpha_{ac}$, $\Delta\alpha_{bc}$, and $F$ together with the control parameters $\tau$ and $r$. To decrease the number of parameters, we empirically determine the value of $F$ that leads to the best degree of alignment for each molecule. To this end, we first choose six molecules (Table I) on the basis of the values of the asymmetry parameter ranging from $\kappa \approx -1$ to $\kappa \approx 0$ at approximately every 0.2, regardless of the value of $\kappa_\alpha$. For each molecule, we calculate $\Phi(\tau, r)$ by systematically changing the total pulse fluence, the time delay $\tau$, and the ratio $r$ to find the best value of $F$. We plot the numerically found six best values of $F$ as a function of $\tilde{B}$, $\tilde{C}$, $\kappa$, $\Delta\alpha_{ac}/\Delta\alpha_{bc}$, $\Delta\alpha_{bc}$, $\kappa_\alpha$, etc., and find the following "empirical" relation

$$F\,[\mathrm{J/cm^2}] = \frac{12.3}{\Delta\alpha_{bc}\,[\mathrm{a.u.}]}, \tag{8}$$

where $F$ and $\Delta\alpha_{bc}$ are measured in units of $\mathrm{J/cm^2}$ and a.u., respectively. When deriving Eq. (8) in Fig. 1, we use the Levenberg–Marquardt algorithm in the nonlinear least squares curve-fitting method. In our trial, we cannot find any clear relationship between $F$ and the other parameters. Because the data corresponding to the six molecules in Fig. 1 do not appear around $\Delta\alpha_{bc} \sim 3$ a.u. and $\Delta\alpha_{bc} \sim 20$ a.u., we employ two more molecules with (g) $\Delta\alpha_{bc} = 3.04$ a.u. and (h) 20.8 a.u. in Table I. We again perform a numerical search for the best values of $F$ through a trial-and-error approach. We see that the two extra results added in Fig.1 [(g) and (h)] almost lie on the fitting curve [Fig. (8)]. For reference, we recalculate the fitting curve by using the eight molecules [(a)-(h)] and the above-mentioned least squares curve-fitting. The

Table I. List of asymmetric top molecules in Fig. 1 [(a)–(f)] used to empirically determine the fluence [Eq. (8)], which is suitably adjusted to achieve the highest degrees of alignment (see text). For each molecule, the value of the asymmetry parameter $\kappa$, that of $\kappa_a$, and that of the degree of alignment (DOA) are also shown. Two extra molecules with (g) $\Delta\alpha_{bc} = 3.04$ a.u. and (h) $\Delta\alpha_{bc} = 20.8$ a.u. are used to illustrate the validity of Eq. (8) (see text).

| molecule | $\kappa$ | $\kappa_a$ | DOA |
|---|---|---|---|
| (a) $C_7H_{12}$ [1)] | -0.06 | -0.24 | 0.64 |
| (b) $SF_4$ | -0.48 | -0.10 | 0.67 |
| (c) $CF_2Cl_2$ | -0.59 | -0.31 | 0.68 |
| (d) $(CH_3)_2CCCH_2$ | -0.64 | -0.76 | 0.73 |
| (e) $CH_3CHCHCH_3$ | -0.84 | -0.22 | 0.68 |
| (f) $SO_2$ | -0.94 | -0.67 | 0.77 |
| (g) $CH_3(CH_2)_5CH_3$ | -0.99 | -0.77 | 0.80 |
| (h) $CH_2CHOCHCH_2$ | -0.99 | -0.26 | 0.68 |

1) Bicyclo[2.2.1]heptane

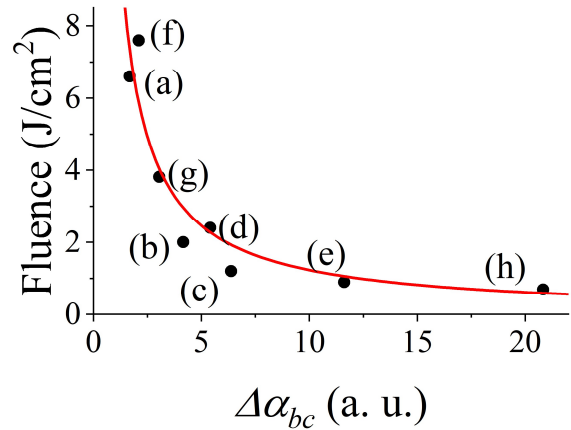

Figure 1

Red curve shows the empirically determined fluence [Eq. (8)] as a function of $\Delta\alpha_{bc}$, which is derived from the Levenberg–Marquardt algorithm with the molecules (a)–(f) in Table I. Fluence associated with extra molecules (g) and (h) in Table I are well fitted by Eq. (8) (see text).

recalculated fitting curve is the same as that in Eq. (8) up to two significant digits in the numerator on the right-hand side of Eq. (8). Because of this, we will use Eq. (8) in the following. By substituting Eq. (8) into Eq. (7), we have the dimensionless interaction expressed as

$$\tilde{V}(t) = -\frac{\eta_0}{4} r \left[ \left(\frac{\Delta\alpha_{ac}}{\Delta\alpha_{bc}}\right)(\boldsymbol{e}_a \cdot \boldsymbol{e}_X)^2 + (\boldsymbol{e}_b \cdot \boldsymbol{e}_X)^2 \right] \varphi_X(t - t_X^0)$$
$$-\frac{\eta_0}{4}(1-r) \left[ \left(\frac{\Delta\alpha_{ac}}{\Delta\alpha_{bc}}\right)(\boldsymbol{e}_a \cdot \boldsymbol{e}_Y)^2 + (\boldsymbol{e}_b \cdot \boldsymbol{e}_Y)^2 \right] \varphi_Y(t - t_Y^0), \quad (9)$$

with the constant $\eta_0 = 24.6/(hA\varepsilon_0 c)$.

## C. Preparing training samples

From Eqs. (5) and (9), we see that the laser-induced rotational dynamics is described by the three molecular parameters $\tilde{B}$, $\tilde{C}$, and $\Delta\alpha_{ac}/\Delta\alpha_{bc}$, that define the coordinate system referred to as a molecular space. Each molecule is uniquely specified by a point in the molecular space, as shown in Fig. 2(a). From the 65 molecules in the molecular space in Fig. 2(a), we first choose 30 training sample molecules by using the Kennard–Stone algorithm [42]. Figure 2(b) shows the projected components on the $\tilde{B}\tilde{C}$-plane, which are distributed within a small region defined by the three lines. Two lines originate from the first criterion, i.e., the asymmetry parameter $-1 < \kappa < 0$ so that $\tilde{C} < \tilde{B}$ and $\tilde{C} > 2\tilde{B} - 1$. The third line originates from the inequality $\tilde{C} > \tilde{B}/(\tilde{B}+1)$, which is derived under the condition $I_{cc} < I_{aa} + I_{bb}$. Within the specified region on the $\tilde{B}\tilde{C}$-plane, the distribution of the points (projected components) is reasonably uniform.

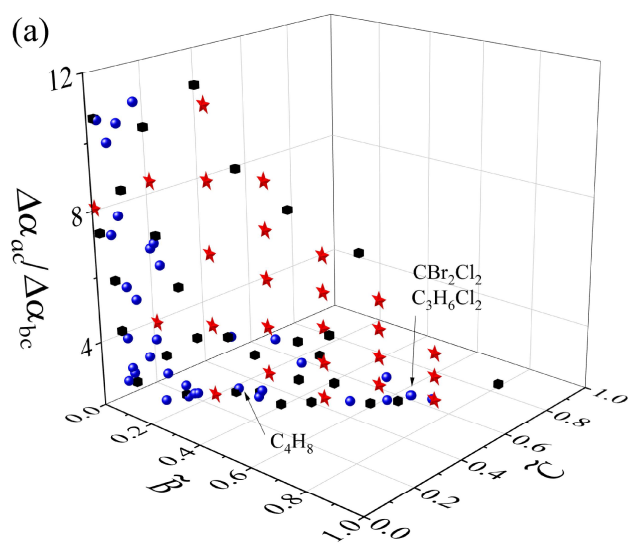

Figure 2(a)

Molecular parameters for the 90 asymmetric top molecules are specified in the present molecular space, the three axes of which represent $\tilde{B}$, $\tilde{C}$, and $\Delta\alpha_{ac}/\Delta\alpha_{bc}$. Among the 55 training sample molecules, 30 (black solid squares) are from the NIST dataset and 25 (red solid stars) are artificial molecules (see text). The blue solid circles correspond to the 35 test sample molecules. The three test sample molecules indicated by arrows with chemical formulas are discussed in Fig. 3.

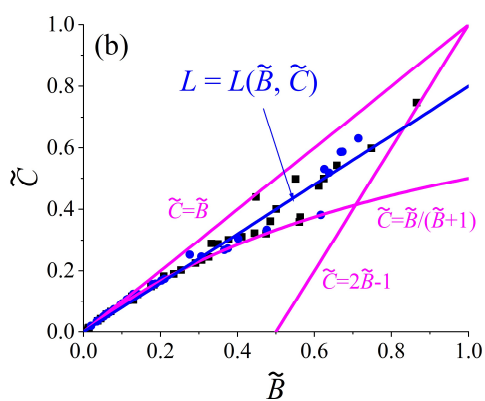

Figure 2(b)

The molecules in Fig. 2(a) are projected onto the $\tilde{B}\tilde{C}$ plane. Black solid squares and blue solid circles correspond to the 30 training and 35 test samples, respectively. The artificial molecules are not shown here. The three purple straight lines that define the projection area are explained in the text. The blue straight line $L = L(\tilde{B}, \tilde{C})$ is the least squares straight line derived from the 65 points.

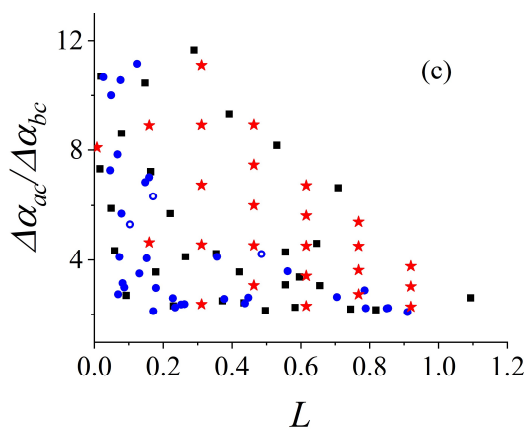

Figure 2(c)

The molecules in Fig. 2(a) are projected onto the plane defined by the two straight lines, $L=L(\tilde{B},\tilde{C})$ [see Fig. 2(b)] and the $\Delta\alpha_{ac}/\Delta\alpha_{bc}$ axis. Red solid stars correspond to the projected 25 artificial molecules. Black solid squares, which correspond to the black solid squares in Fig. 2(a), are the 30 training sample molecules. The test sample molecules are shown by the 3 blue open circles and the 32 blue solid circles, the former of which specify the molecules used to evaluate the generic error (see text).

On the other hand, the points projected onto the other two planes defined by the $\Delta\alpha_{ac}/\Delta\alpha_{bc}$ axis exhibit a non-uniform distribution (not shown). We see from Fig. 2(a) that the 30 training samples are localized in specific regions in the molecular space, making it difficult to train the CNN model. To facilitate the CNN model training, we take advantage of the fact that the rotational dynamics is specified by a small number of molecular parameters $\tilde{B}$, $\tilde{C}$, and $\Delta\alpha_{ac}/\Delta\alpha_{bc}$. Because of this, we can introduce "artificial" molecules to realize a reasonably uniform distribution of the training samples in the molecular space so that the test samples, i.e., the other molecules not included in the training set, are reasonably interpolated. We adopt the following procedure. First, we introduce a rectangle with $L=L(\tilde{B},\tilde{C})$ being one of the sides on the $\tilde{B}\tilde{C}$-plane. It is the least squares straight line derived by using the 65 molecules [Fig. 2(b)]. To effectively introduce the artificial molecules, the other side of the rectangle is set parallel to the $\Delta\alpha_{ac}/\Delta\alpha_{bc}$ axis scaled by $\tilde{B}+\tilde{C}$ because more (less) molecules are originally distributed when the value of $\tilde{B}+\tilde{C}$ is small (large) in Fig. 2(a). We then divide the rectangle into $7\times7$ small rectangles to introduce 64 lattice points into the molecular space. We intuitively extract the 25 lattice points that correspond to the artificial molecules, which are added to the training sample molecules, as shown in Fig. 2(c). The 55 training samples are composed of the originally chosen 30 molecules and the 25 artificial molecules, which are shown by the black solid squares and the red solid stars, respectively, in Fig. 2(a).

We already introduced the dimensionless time [Eq. (5)] to treat various molecules in a unified manner. Even so, the values of the dimensionless rotational constants $\tilde{B}$ and $\tilde{C}$, which characterize a "slow" rotational motion, are still distributed over an approximately single-digit range in Fig. 2. This feature introduces markedly different temporal behavior into the rotational dynamics depending on the molecule, leading to difficulty in control landscape map prediction by the CNN model. To solve this problem, we measure the dimensionless time in units of $T_{\rm rot} = \pi/(\tilde{B}+\tilde{C})$, which will be referred to as the rotational period for convenience. As will be discussed in Sec. III, $T_{\rm rot}$ measures the slow dynamics in a unified way probably because of $\pi/(2\tilde{B}) < T_{\rm rot} < \pi/(2\tilde{C})$. By using $T_{\rm rot}$, the normalized intensity envelope functions are defined by

$$\varphi_X(t-t_X^0) = \frac{1}{\sqrt{2\pi}\sigma}\exp\left[-\frac{(t-t_X^0)^2}{\sigma^2}\right] \text{ and } \varphi_Y(t-t_Y^0) = \frac{1}{\sqrt{2\pi}\sigma}\exp\left[-\frac{(t-t_Y^0)^2}{\sigma^2}\right], \quad (10)$$

where the FWHM ($2\sqrt{\ln 2}\sigma$) and $t_X^0$ are set to $0.003T_{\rm rot}$ and $0.1T_{\rm rot}$, respectively.

The control parameters are uniformly discretized such that the time delay $\tau_i = (i-1)\Delta\tau$ ($i=1,2,\cdots,120$) with $\Delta\tau = 0.01T_{\rm rot}$ ($\gg 0.003T_{\rm rot}$) and the fluence ratio $r_j = (j-1)\Delta r$ ($j = 1, 2, \cdots, 50$) with $\Delta r = 0.01$ so that the control landscape map of each molecule consists of 6000 pixels. We emphasize again that each pixel in the single landscape map represents the maximum degree of alignment $\Phi(\tau_i, r_j)$ under the given set of control parameters $\tau_i$ and $r_j$. We search the maximum degree of alignment $\Phi(\tau_i, r_j)$ by assuming $t_0 = 0$ and $t_f = t_Y^0 + 1.2 T_{rot}$. We thus have to numerically solve Eqs. (5) and (6) 6000 times even to prepare a single control landscape map, i.e., a single training sample. To circumvent this high computational cost, we utilize the CNN model as explained below. For each landscape map of the training sample molecule, we randomly choose 600 sets of control parameters and calculate 600 $\{\Phi(\tau_i, r_j)\}$, which are regarded as the training samples to train the CNN model to generate a total of 6000 values. We adopt and train the CNN model for each training sample molecule within the Keras framework [43]. According to Ref. [37], each set of control parameters $\tau_i$ and $r_j$ is represented by the sparse $120 \times 50$ matrix, that is, one of the matrix elements that correspond to the specified time delay $\tau_i$ ($i = 1, 2, \cdots, 120$), and the fluence ratio $r_j$ ($j = 1, 2, \cdots, 50$) is set to 1, while all the other elements are set to 0. This input matrix and its label $\Phi(\tau_i, r_j)$ consist of one of the training samples for the CNN model. The input matrix is sent to the two convolutional layers composed of four $5 \times 5$ filters and sixteen $5 \times 5$ filters, respectively; to the single fully connected layer with the 64 units; and finally to the output layer with a single unit. The output layer provides the predicted value $\Phi^{(pred)}(\tau_i, r_j)$. We use the ReLU activation function for all layers except the output layer in which the sigmoid function is used. We finish training the CNN model when the value of MSE associated with the 600 sets of $\{\Phi(\tau_i, r_j) \text{ and } \Phi^{(pred)}(\tau_i, r_j)\}$ is smaller than $10^{-6}$. We repeat the procedure for all the training sample molecules to prepare the training set. In Appendix B, the two typical examples of the scattered plots associated with the training samples are provided to illustrate the validity of the above-mentioned procedure.

**D. Training CNN model**

We develop the CNN model to predict the landscape maps of the 35 test sample molecules, which are not included in the training set (Sec. IIC). The input matrix that represents the control parameters is sent to the two convolutional layers, both of which are composed of four $5\times 5$ filters, and then to the two fully connected NN layers with 300 and 50 units. On the other hand, the inputs associated with the molecular parameters $\tilde{B}$, $\tilde{C}$, and $(\Delta\alpha_{ac}/\Delta\alpha_{bc})/(\Delta\alpha_{ac}/\Delta\alpha_{bc})_{max}$ with $(\Delta\alpha_{ac}/\Delta\alpha_{bc})_{max}=11.7$ are sent to the other two fully connected NN layers with 50 and 1000 units. Here, $(\Delta\alpha_{ac}/\Delta\alpha_{bc})_{max}$ is introduced to normalize the value of $\Delta\alpha_{ac}/\Delta\alpha_{bc}$. The outputs of the 1000 units are simply added to those of the 50 units associated with the control parameters to form the layer with 1050 units, which are sequentially sent to the fully connected layers with 100, 50, and 11 units, and finally to the output layer. We always use the ReLU activation function for all layers except the output layer. The output layer with the sigmoid activation function provides the predicted value $\Phi^{(\text{pred})}(\tau_i, r_j)$ for the input molecule under the specified control parameters $\tau_i$ and $r_j$. The CNN model is trained to minimize the mean squared error (MSE) defined by

$$\Delta = \frac{1}{N}\sum_{n=1}^{N}\frac{1}{120\times 50}\sum_{i=1}^{120}\sum_{j=1}^{50}\left|\Phi_n^{(\text{pred})}(\tau_i, r_j)-\Phi_n(\tau_i, r_j)\right|^2, \tag{11}$$

where the subscript "$n$" in $\Phi_n^{(\text{pred})}$ and $\Phi_n$ is introduced to distinguish the $N=55$ training sample molecules. We estimate the generic error by using the 31st, 32nd, and 33rd test sample molecules chosen by the Kennard–Stone algorithm in the molecular space. We use the same expression as that in Eq. (11) for the three molecules to evaluate the generic error. To avoid overlearning, we regard the generic error of $4.8 \times 10^{-4}$ as the converged value. When the convergence is achieved, the MSE in Eq. (11) has a value of $\Delta = 2.5 \times 10^{-4}$.

## III. Results and discussion

We apply the trained CNN model to predict the control landscape maps of the 35 test sample molecules that are not involved in the training set. The control landscape map of each molecule shows the maximum degrees of alignment $\Phi^{(\text{pred})}(\tau_i, r_j)$ as a function of the discretized $\tau_i$ and $r_j$. In each control landscape map, the "maximum" value is referred to as the optimal value because it is the optimal degree of 3D alignment corresponding to the optimal control solution within the present double pulse control. Here, the predicted optimal value and the numerically obtained (true) optimal value are denoted by $\Phi_{\text{opt}}^{(\text{pred})}$ and $\Phi_{\text{opt}}^{(\text{true})}$, respectively, which also specify their optimal time delays and optimal ratios.

It is apparent that one of the direct applications of the landscape maps is to predict the optimal value $\Phi_{\text{opt}}^{(\text{pred})}$ for each molecule. In Fig. 3, we show the optimal values associated with the 35 test sample molecules in descending order. We also show the numerically obtained (true) optimal values using the purple solid triangles, which are obtained as follows. In each predicted landscape map composed of 6000 pixels, we focus on the top 5%, i.e., the 300 pixels according to their degrees of alignment, and sort them in descending order in value. We first select the

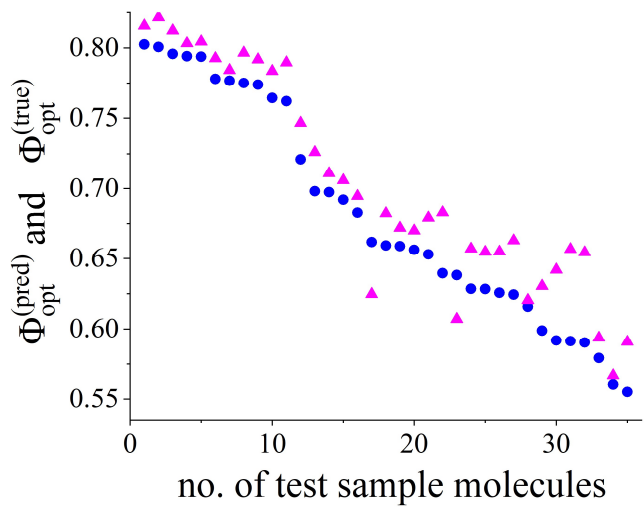

Figure 3

For the 35 test sample molecules listed in Appendix C (Table II), the blue solid circles show the predicted optimal values $\{\Phi_{opt}^{(pred)}\}$ in descending order. The purple solid triangles show the numerically obtained (true) optimal values $\{\Phi_{opt}^{(true)}\}$ (see text).

pixel with the largest value and regard it as one of the candidates for the true optimal value. Next, we select the pixel with the second largest value. This pixel is selected as a candidate if it is at least 5 Chebyshev distance (chessboard distance) apart from the previously selected (first) pixel. If not, we simply ignore the second pixel. By repeating the procedure 300 times, we prepare a set of candidate pixels associated with the true optimal value. For each candidate pixel, we define the $5\times 5$ pixel region, the center of which is the candidate pixel. Then, we numerically calculate the degrees of alignment for all the pixel regions to find the "maximum" value, which is regarded as the true optimal value for each molecule. In Fig. 3, we see a good agreement between the predicted $\Phi_{opt}^{(pred)}$ and the true $\Phi_{opt}^{(true)}$ optimal values especially when the optimal values are larger than ca. 0.65. For values smaller than ca. 0.65, the CNN model still reasonably predicts the optimal values although some of the predicted optimal values are slightly different from those obtained numerically. The difficulty in predicting the optimal values of some molecules can be partly attributed to their near-edge positions in the molecular space [Fig. 2(a)].

We see from Fig. 3 that the 35 optimal values are widely distributed from ca. 0.55 to ca. 0.8 even though we have restricted ourselves to the prolate-type $C_{2v}$ asymmetric top molecules. We expect that the differences in the optimal values can be interpreted in terms of the asymmetry parameter $\kappa$ and $\Delta\alpha_{ac}/\Delta\alpha_{bc}$ that characterize the free rotational motion and the laser-induced interaction, respectively. In Fig. 4, we plot $\Phi_{\text{opt}}^{(\text{pred})}$ of the 87 molecules including the 25 artificial molecules as a function of $\kappa$ and $\Delta\alpha_{ac}/\Delta\alpha_{bc}$. Note that we solely use the predicted optimal values $\Phi_{\text{opt}}^{(\text{pred})}$ and ignore the three molecules numbered 30, 31, and 32 in Fig. 3 and Table II (Appendix C) because of the large deviations in predicted value ($>0.05$). Roughly speaking, we see a curved distribution from top to bottom in Fig. 4. To examine the distribution pattern in detail, we apply the hierarchical clustering approach to classify the 87 optimal values into the three clusters. First, we normalize the predicted landscape map of each molecule with respect to its optimal value and regard the 6000 points in each normalized landscape map as the 6000-dimensional vectors. By using the 6000-dimensional vectors associated with the control landscape maps, we calculate the Euclidean distances between them. On the basis of the Euclidean distances, we apply the agglomerative approach [44] together with Ward's method [45], a hierarchical clustering approach, to classify the landscape maps into the three clusters named Clusters 1, 2, and 3. In Fig. 4, the optimal values belonging to Clusters 1, 2, and 3 are specified by red, blue, and black solid colors, respectively. The molecules in Cluster 1 have large optimal values and are characterized by large values of $|\kappa|>0.65$ and $\Delta\alpha_{ac}/\Delta\alpha_{bc}>5$. The molecules in Cluster 3 do not have large optimal values and are characterized by the small values of $\Delta\alpha_{ac}/\Delta\alpha_{bc}<4.5$, while being widely distributed along the $\kappa$ axis. The molecules in Cluster 2 have small values of $3<\Delta\alpha_{ac}/\Delta\alpha_{bc}<4.5$ and large values of $|\kappa|>0.85$, indicating that these molecules possess characteristics intermediate of those of Clusters 1 and 3.

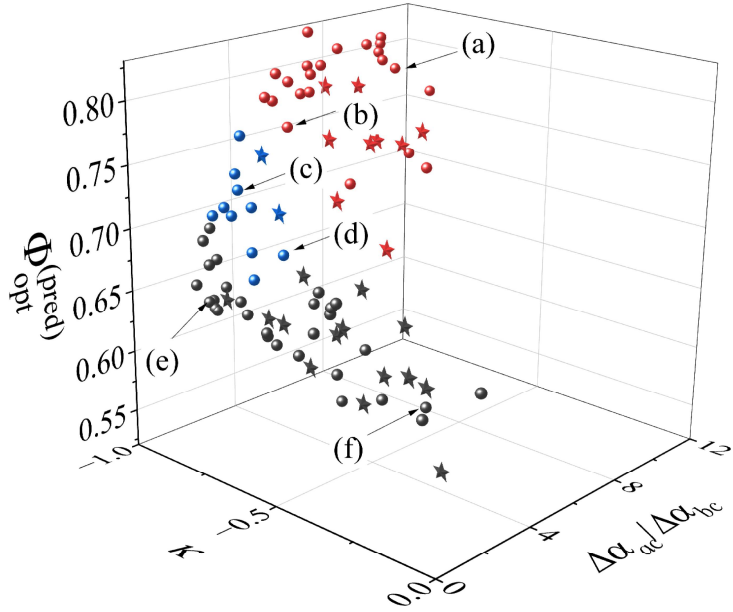

Figure 4

Predicted optimal degrees of alignment $\Phi_{opt}^{(pred)}$ with respect to the asymmetry parameter $\kappa$ and $\Delta\alpha_{ac}/\Delta\alpha_{bc}$. Red, blue, and black solid circles (stars) correspond to the 62 molecules (25 artificial molecules) classified into Clusters 1, 2, and 3, respectively (see text).

Finally, we consider the control mechanisms in each cluster through the typical examples of the control landscape maps. Figures 5(a) and (b) show that the typical landscape maps belonging to Cluster 1, which are characterized by the two sharp peaks around $\tau \sim 0.4T_{rot}$ and $\tau \sim 1.0T_{rot}$ with the small values of $r$. Here, the small values of $r$ mean that the fluence of the first *X*-polarized pulse is (much) weaker than that of the second *Y*-polarized pulse. The 3D alignment control mechanisms for Cluster-1 molecules can be understood according to Refs. [26] and [27]. Because the value of $\Delta\alpha_{ac}$ is considerably larger than that of $\Delta\alpha_{bc}$ in Cluster-1 molecules, the first pulse with weak pulse fluence selectively and highly aligns the molecular *a*-axis along the *X*-axis (1D alignment). Then, the second strong pulse polarized along the *Y*-axis kicks the molecule at the timings when the *a*-axis maximally 1D aligns along the *X*-axis to minimize the torque exerted on the *a*-axis.

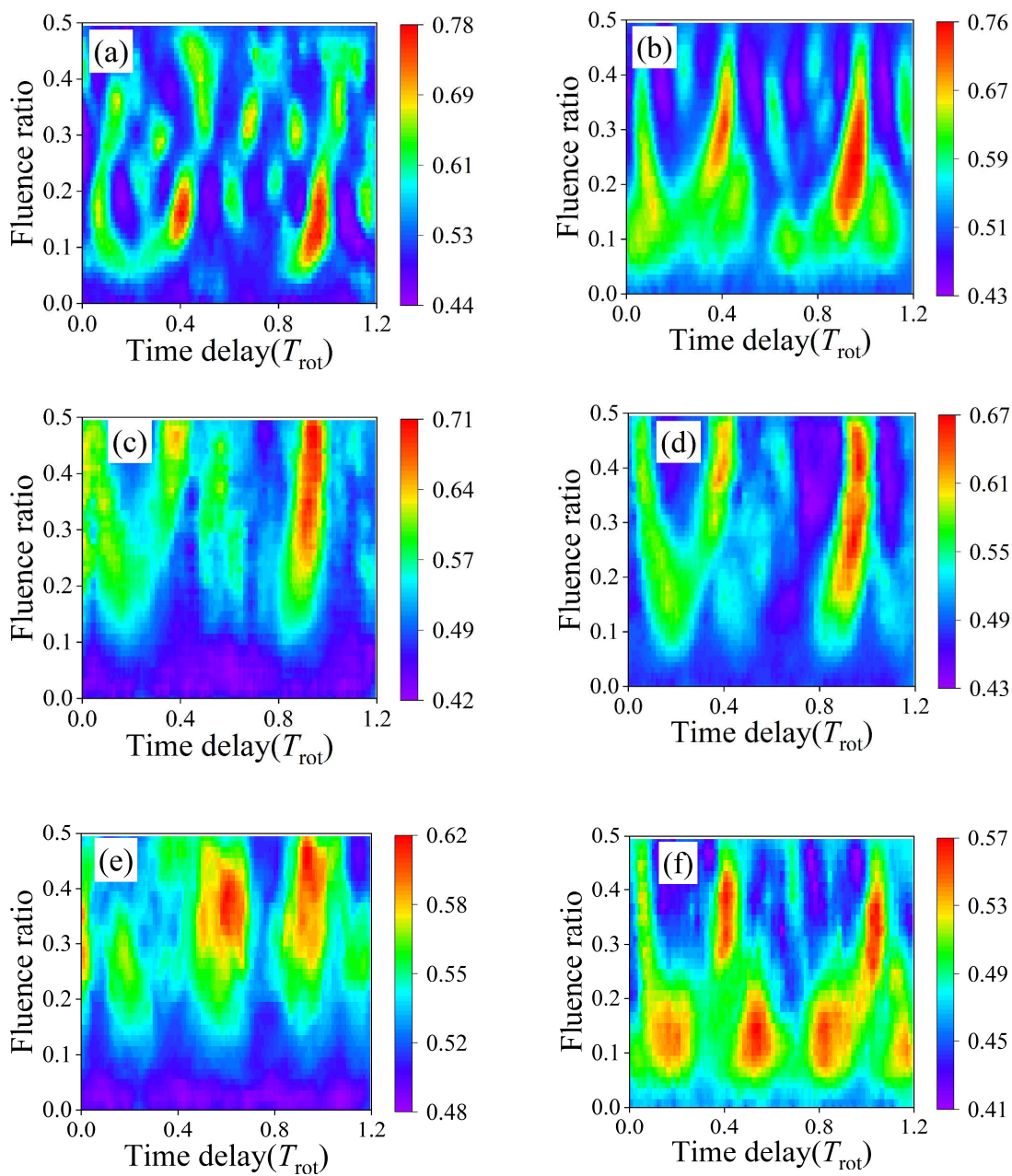

Figure 5

Typical examples of control landscape maps of the asymmetric top molecules classified into Cluster 1 [(a) and (b)], Cluster 2 [(c) and (d)], and Cluster 3 [(e) and (f)] are shown. We use molecule-dependent color scales to clearly show the distribution patterns. (a) $CH_2BH$ ( $\kappa = -0.98$, $\Delta\alpha_{ac}/\Delta\alpha_{bc} = 11.2$ ), (b) $SO_2$ ( $-0.94$, $5.70$ ), (c) $CH_2Cl_2$ ( $-0.98$, $3.50$ ), (d) $SeO_2$ ( $-0.83$, $4.23$ ), (e) $C_4H_2O_3$ ( $-0.73$, $2.40$ ), and (f) $BeCO_3$ ( $-0.37$, $4.61$ ). Molecules (a), (c), and (e) belong to the test set, and molecules (b), (d), and (f) belong to the training set. We use molecule-dependent color scales to clearly show the distribution patterns.

Examples of the control landscape maps belonging to Cluster 2 are shown in Figs. 5(c) and (d), which are characterized by the single narrow peaks around $\tau \sim 1.0 T_{\text{rot}}$ along the $\tau$-axis while the peaks are widely distributed along the $r$-axis. We can understand the features of the Cluster-2 molecules in terms of the large values of $|\kappa| > 0.85$, meaning that the Cluster-2 molecules are similar in shape to the prolate symmetric top molecules ($\kappa = -1$). The rotational wave packets of the Cluster-2 molecules show almost regular and clear revivals after the first pulse, which determines the best timing of the second pulse $\sim 1.0 T_{\text{rot}}$. The small values of $3 < \Delta\alpha_{ac}/\Delta\alpha_{bc} < 4.5$, on the other hand, reduce the degrees of 1D alignment of the $a$-axis along the $X$-axis by the first pulse. Because of this, the $a$-axis is slightly affected by the torque exerted by the second pulse, resulting in a slight decrease in the degree of 3D alignment relative to that of the Cluster-1 molecules.

Regarding the Cluster-3 molecules, the small values of $\Delta\alpha_{ac}/\Delta\alpha_{bc}$ make it difficult to selectively 1D align the most polarizable molecular $a$-axes along the $X$-axis, leading not only to a further reduction of the degree of 3D alignment but also to a variety of structures of the landscape maps depending on their molecular parameters, as shown in Figs. 5(e) and (f). For Cluster-3 molecules, we may need to shape the laser pulses beyond the double pulse excitation schemes considered here in order to achieve higher degrees of 3D alignment.

## Summary

We have demonstrated the effectiveness of the CNN-based approach for predicting the control landscape maps of the full quantum molecular dynamics through a case study of the laser-induced 3D alignment of asymmetric top molecules with the double pulse control scheme. The control landscape map of each molecule is defined by the maximum 3D degree of alignment $\Phi(\tau_i, r_j)$ as a function of the discretized $\tau_i$ ($i = 1, 2, \cdots, 120$) and $r_j$ ($j = 1, 2, \cdots, 50$) with $\tau$ and $r$ being the time delay and the fluence ratio, respectively. We have restricted ourselves to the prolate-type asymmetric top molecules with the $C_{2v}$ symmetry to prepare the training samples at a reasonable computational cost. We have assumed the low-temperature limiting case as the most remarkable quantum behavior of the rotational wave packets is expected. By using the empirically derived fluence formula [Eq. (8)], the molecules are shown to be specified in the molecular space defined by the three axes, i.e., the two dimensionless rotational constants $\tilde{B}$ and $\tilde{C}$, and the ratio of the polarizabilities $\Delta\alpha_{ac}/\Delta\alpha_{bc}$. To realize the uniform distribution of the training sample molecules in the molecular space, we have introduced the artificial molecules by taking advantage of the fact that the rotational dynamics is specified by the rotational constants and the polarizability components. This is practically important because the CNN model is known to be especially good at data interpolation.

The present CNN model is trained by 55 landscape maps of the training sample molecules including the 25 artificial molecules. The trained CNN models have successfully predicted the control landscape maps of the test sample molecules with reasonably high accuracy as illustrated by the screening for the optimal degrees of 3D alignment. As the optimal values are widely distributed from ca. 0.55 to ca. 0.8 (Fig. 3), we have applied the agglomerative approach to classify all the landscape maps, i.e., all the molecules, into the three groups. The control landscape maps belonging to each cluster have common structural characteristics, reflecting the effectiveness of the 3D alignment control. The double pulse control is especially effective for the molecules that have a large value of $\Delta\alpha_{ac}/\Delta\alpha_{bc}$, which would be in good agreement with the control mechanism proposed in Refs. [26] and [27].

The successfully predicted control landscape maps in the present case study strongly suggest the usefulness of applying the ML approaches to the laser control of full quantum molecular dynamics. When considering the quantum control of molecular dynamics, it is natural to investigate the optimal methods, e.g., based on optimal control theory. As the optimal control approaches actively cooperate with the complicated quantum interferences among the molecular wave functions, the control mechanisms are often too complicated to be understood by our chemical physics-based intuitions and/or considerations. In this regard, quantum optimal control approaches in combination with ML-based interpretations such as control landscape maps would be useful not only for quantum molecular dynamics but also for other quantum technologies.

**Acknowledgements**

YO acknowledges support in the form of a Grant-in-Aid for Scientific Research (C) (23K04659) and partial support from the Joint Usage/Research Program on Zero-Emission Energy Research, Institute of Advanced Energy, Kyoto University (ZE2024B-17). This work was also supported by JST SPRING, Grant Number JPMJSP2114. Computations were performed in the Research Center for Computational Science, Okazaki, Japan (Project: 22-IMS-C987).

**APPENDIX A: Time evolution of rotational wave packets**

We briefly summarize the numerical method to solve the dimensionless Schrödinger equation in Eq. (5). We expand the rotational wave packet in terms of the eigenfunctions of the symmetric top $\{|JKM\rangle\}$ so that we have

$$|\psi(t)\rangle = \sum_{J=0}^{J_{\max}} \sum_{K,M=-J}^{J} C_{KM}^{J}(t)|JKM\rangle \ . \tag{A1}$$

Here, the eigenfunction $|JKM\rangle$ is also expressed as the complex conjugate of the Wigner rotational matrix [38]

$$\langle \hat{R}|JKM\rangle = \sqrt{\frac{2J+1}{8\pi^2}} D_{MK}^{J*}(\hat{R}), \tag{A2}$$

where the set of Euler angles is denoted as $\hat{R}$. If we substitute Eq. (A1) into Eq. (5), we have the equations of motion for $\{C_{KM}^J(t)\}$,

$$i\dot{C}_{KM}^J(t) = \sum_{J'M'K'} \langle JKM|[\tilde{H}_0 + \tilde{V}(t)]|J'K'M'\rangle C_{K'M'}^{J'}(t), \tag{A3}$$

which is expressed in terms of the matrix elements of the dimensionless field-free Hamiltonian $\tilde{H}_0$ and the dimensionless interaction potential $\tilde{V}(t)$ [Eq. (9)]. We rewrite $\tilde{H}_0$ as

$$\tilde{H}_0 = J_a^2 + \tilde{B}J_b^2 + \tilde{C}J_c^2 = \left(\frac{1+\tilde{B}}{2}\right)\boldsymbol{J}^2 + \left(\tilde{C} - \frac{1+\tilde{B}}{2}\right)J_c^2 + \left(\frac{1-\tilde{B}}{4}\right)(J_+^2 + J_-^2). \tag{A4}$$

Here, the operators $J_\pm = J_a \pm iJ_b$ have the following non-zero matrix elements [38]

$$\langle JK \mp 2M|J_\pm^2|JKM\rangle = \sqrt{J(J+1) - K(K \mp 1)}\sqrt{J(J+1) - (K \mp 1)(K \mp 2)}. \tag{A5}$$

We also need to calculate the matrix elements associated with the squares of the direction cosines such as $(\boldsymbol{e}_a \cdot \boldsymbol{e}_X)^2$, $(\boldsymbol{e}_a \cdot \boldsymbol{e}_Y)^2$, $\cdots$, and $(\boldsymbol{e}_c \cdot \boldsymbol{e}_Z)^2$, which are involved in $\tilde{V}(t)$ and in the operators associated with the degrees of alignment in Eq. (6). With the aid of the Clebsch-Gordan series [38], we have

$$(e_a \cdot e_X)^2 = \frac{1}{12}\left[4D_{00}^0 + 2D_{00}^2 - \sqrt{6}\left(D_{20}^2 + D_{-20}^2 + D_{02}^2 + D_{0-2}^2\right) + 3\left(D_{22}^2 + D_{-22}^2 + D_{2-2}^2 + D_{-2-2}^2\right)\right] \quad \text{(A6)}$$

$\cdots$, and

$$(e_c \cdot e_Z)^2 = \frac{1}{3}\left(D_{00}^0 + 2D_{00}^2\right), \quad \text{(A7)}$$

where we omit $\hat{R}$ in $\{D_{MK}^J(\hat{R})\}$ for simplicity. All the matrix elements are reduced to the integrals over the triple products of the Wigner rotation matrices [38], which are expressed in terms of the Clebsch-Gordan coefficients. We substitute the matrix elements into the equations of motion for $\{C_{KM}^J(t)\}$ and numerically integrate Eq. (A3) by the 5th-order Runge–Kutta method.

**APPENDIX B: Preparing control landscape maps of training sample molecules**

The control landscape map of each molecule is composed of 6000 pixels that represent $\{\Phi(\tau_i, r_j): i = 1, 2, \cdots, 120 \text{ and } j = 1, 2, \cdots, 50\}$. For each training sample molecule, as explained in Sec. IIC, we numerically calculate the randomly chosen 600 $\{\Phi(\tau_i, r_j)\}$ and use them to predict the values of the 6000 pixels. We evaluate the validity of this procedure by considering the randomly chosen two training sample molecules, CH$_2$CS (Fig. 6) and CH$_3$CSCH$_3$

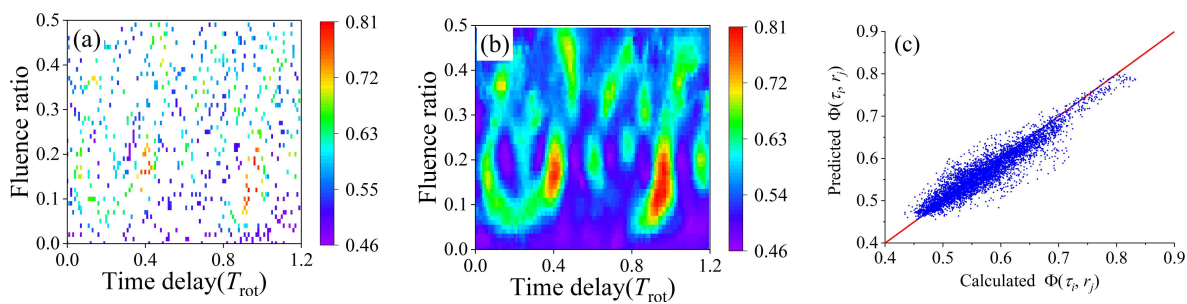

Figure 6

Landscape map and scattered plot associated with the training sample molecule $CH_2CS$. (a) Initially chosen (numerically calculated) 600 $\{\Phi(\tau_i, r_j)\}$, (b) the predicted landscape map (see text), and (c) the scatter plot of all the 6000 $\{\Phi(\tau_i, r_j)\}$.

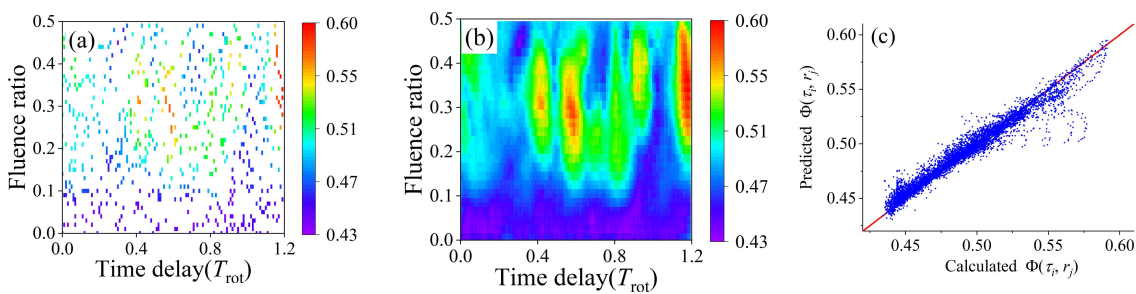

Figure 7

Landscape map and scattered plot associated with the training sample molecule $CH_3CSCH_3$. (a) Initially chosen (numerically calculated) 600 $\{\Phi(\tau_i, r_j)\}$, (b) the predicted landscape map (see text), and (c) the scatter plot of all the 6000 $\{\Phi(\tau_i, r_j)\}$.

(Fig. 7). From the initially prepared (numerically calculated) 600 $\{\Phi(\tau_i, r_j)\}$ in Figs. 6(a) and 7(a), the CNN model constructs the control landscape maps in Figs. 6(b) and 7(b) with reasonably high probability. This is confirmed by the scatter plots in Figs. 6(c) and 7(c), both of which are composed of the 6000 values of $\{\Phi(\tau_i, r_j)\}$.

# APPENDIX C: List of test sample molecules in Fig. 3

Table II provides the test sample molecules, the optimal degrees of 3D alignment of which are shown in Fig. 3. For convenience, we number the test sample molecules in descending order of the value of $\Phi_{\text{opt}}^{(\text{pred})}$ such that $\Phi_{n,\text{opt}}^{(\text{pred})}$ ($n = 1, 2, \cdots, 35$).

Table II. List of 35 test sample molecules in Fig. 3, numbered in descending order of the value of $\Phi_{\text{opt}}^{(\text{pred})}$. The dimensionless rotational constants $\tilde{B}$ and $\tilde{C}$, $\kappa$, $\Delta\alpha_{ac}/\Delta\alpha_{bc}$, and cluster classification numbers (#cluster), which are examined in Fig. 4, are also listed.

| no. | molecule | $\tilde{B}$ | $\tilde{C}$ | $\kappa$ | $\Delta\alpha_{ac}/\Delta\alpha_{bc}$ | #cluster |
|---|---|---|---|---|---|---|
| 1 | $CH_3(CH_2)_7CH_3$ | 0.037 | 0.036 | -0.999 | 10.0 | 1 |
| 2 | $CH_2CS$ | 0.020 | 0.019 | -0.999 | 10.7 | 1 |
| 3 | $BrOBr$ | 0.035 | 0.034 | -0.998 | 7.27 | 1 |
| 4 | $ONONO$ | 0.051 | 0.049 | -0.995 | 7.84 | 1 |
| 5 | $C_3H_7OC_3H_7$ | 0.058 | 0.056 | -0.997 | 10.6 | 1 |
| 6 | $CH_2BH$ | 0.096 | 0.088 | -0.982 | 11.2 | 1 |
| 7 | $S_3$ | 0.125 | 0.111 | -0.969 | 7.02 | 1 |
| 8 | $CH_3(CH_2)_3CH_3$ | 0.112 | 0.107 | -0.987 | 6.85 | 1 |
| 9 | $H_2CS$ | 0.060 | 0.057 | -0.993 | 5.69 | 1 |
| 10 | $Cl_2O$ | 0.080 | 0.074 | -0.987 | 5.30 | 1 |
| 11 | $B_2H_4$ | 0.129 | 0.125 | -0.989 | 6.33 | 1 |
| 12 | $CH_2ClCH_2CH_2Cl$ | 0.055 | 0.053 | -0.996 | 4.11 | 2 |
| 13 | $C_2H_5SC_2H_5$ | 0.117 | 0.109 | -0.982 | 4.06 | 2 |
| 14 | $CH_2Cl_2$ | 0.101 | 0.093 | -0.983 | 3.50 | 2 |
| 15 | $HS(CH_2)_3SH$ | 0.062 | 0.060 | -0.994 | 3.15 | 2 |
| 16 | $BHCl_2$ | 0.067 | 0.063 | -0.991 | 2.99 | 3 |
| 17 | $CF_2Cl_2$ | 0.626 | 0.530 | -0.595 | 2.88 | 3 |
| 18 | $C_6H_{12}$[1)] | 0.276 | 0.252 | -0.936 | 4.11 | 2 |
| 19 | $NCCH_2CN$ | 0.140 | 0.125 | -0.965 | 2.97 | 3 |
| 20 | $SeF_4$ | 0.638 | 0.518 | -0.503 | 2.23 | 3 |
| 21 | $CHOOCHO$ | 0.053 | 0.050 | -0.994 | 2.74 | 3 |
| 22 | $N_2H_2$ | 0.134 | 0.118 | -0.964 | 2.13 | 3 |
| 23 | $SiCl_2(CH_3)_2$ | 0.714 | 0.631 | -0.547 | 2.09 | 3 |
| 24 | $SiH_2Cl_2$ | 0.180 | 0.157 | -0.944 | 2.60 | 3 |
| 25 | $COBr_2$ | 0.187 | 0.158 | -0.930 | 2.26 | 3 |
| 26 | $SCl_2$ | 0.202 | 0.168 | -0.918 | 2.37 | 3 |
| 27 | $C_8H_6$[2)] | 0.211 | 0.174 | -0.911 | 2.38 | 3 |
| 28 | $C_4H_2O_3$ | 0.366 | 0.268 | -0.732 | 2.41 | 3 |
| 29 | $CH_3NNCH_3$ | 0.402 | 0.305 | -0.721 | 4.22 | 3 |
| 30 | $CH_3CHCHCH_3$ | 0.307 | 0.246 | -0.840 | 2.57 | 3 |
| 31 | $CH_3CCl_2CH_3$ | 0.669 | 0.587 | -0.603 | 2.22 | 3 |
| 32 | $CBr_2Cl_2$ | 0.673 | 0.587 | -0.585 | 2.24 | 3 |
| 33 | $BrF_3$ | 0.375 | 0.273 | -0.719 | 2.62 | 3 |
| 34 | $C_2H_4CO_3$ | 0.476 | 0.334 | -0.572 | 3.58 | 3 |
| 35 | $Li_2CO_3$ | 0.617 | 0.381 | -0.239 | 2.64 | 3 |

1) cis-1,3-dimethylcyclobutane

2) 5-(2-Cyclopropen-1-ylidene)-1,3-cyclopentadiene